\documentclass[11pt]{article}

\usepackage[T1]{fontenc}
\usepackage[utf8]{inputenc}

\usepackage{fancyhdr} 

\usepackage[inline]{enumitem}
\usepackage{multicol,multirow,booktabs}

\usepackage{times}

\usepackage[hang,small,labelfont=bf,up,textfont=it,up]{caption}
\usepackage{subcaption}
\usepackage{amsmath,amsfonts,amsthm,amssymb,bm,dsfont}	
\setlist{nolistsep} 

\usepackage{graphicx}
\usepackage[a4paper,top=2.5cm,bottom=2.5cm,left=2.2cm,right=2.2cm]{geometry}

\usepackage{numprint}
\npthousandsep{,}\npthousandthpartsep{}\npdecimalsign{.}

\usepackage{tikzit}
\usepackage{pgfplots}
\usetikzlibrary{backgrounds, fit, positioning}

\tikzstyle{node}=[fill=none, draw=black, shape=circle, tikzit category=Graphs, tikzit fill=white, minimum size=1cm]
\tikzstyle{rednode}=[fill=none, draw={rgb,255: red,208; green,0; blue,0}, shape=circle, tikzit category=Graphs, tikzit fill=white, minimum size=1cm]
\tikzstyle{greennode}=[fill=none, draw={rgb,255: red,0; green,143; blue,0}, shape=circle, tikzit category=Graphs, tikzit fill=white, minimum size=1cm]
\tikzstyle{thicknode}=[fill=none, draw=black, shape=circle, thick, tikzit fill=white, minimum size=1cm]
\tikzstyle{rectangle}=[fill=none, draw=black, shape=rectangle, tikzit fill=white, minimum width=0.5cm, minimum height=0.75cm]
\tikzstyle{square}=[fill=none, draw=black, shape=rectangle, tikzit fill=white, minimum width=0.75cm, minimum height=0.75cm]
\tikzstyle{redrectangle}=[fill=none, draw={rgb,255: red,208; green,0; blue,0}, shape=rectangle, tikzit fill=white, minimum width=0.5cm, minimum height=0.75cm]
\tikzstyle{greenrectangle}=[fill=none, draw={rgb,255: red,0; green,143; blue,0}, shape=rectangle, tikzit fill=white, minimum width=0.5cm, minimum height=0.75cm]
\tikzstyle{abar}=[fill={rgb,255: red,191; green,191; blue,191}, draw=black, shape=rectangle, minimum width=0.25cm, minimum height=0.35cm]
\tikzstyle{bbar}=[fill={rgb,255: red,208; green,0; blue,0}, draw=black, shape=rectangle, minimum width=0.25cm, minimum height=0.75cm]
\tikzstyle{cbar}=[fill={rgb,255: red,191; green,191; blue,191}, draw=black, shape=rectangle, minimum width=0.25cm, minimum height=0.25cm]
\tikzstyle{smallnode}=[fill=none, draw=black, shape=circle, tikzit fill=white, minimum size=0.6cm]

\tikzstyle{edge}=[fill=none, draw=black, -]
\tikzstyle{greenedge}=[-, draw={rgb,255: red,0; green,143; blue,0}, thick]
\tikzstyle{arrow}=[->, thick]
\tikzstyle{thinarrow}=[->]

\pgfplotsset{compat=1.18}

\usepackage{algorithm}
\usepackage[noend]{algorithmic}

\usepackage{caption}
\usepackage{mathdots}
\usepackage{array}
\usepackage{multirow}
\usepackage{tabularx}
\usepackage{booktabs}

\newcommand{\titledoc}{Approximate learning of parsimonious\\Bayesian context trees}
\newcommand{\titleshort}{Approximate learning of parsimonious Bayesian context trees}

\providecommand{\keywords}[1]{{\small{\textbf{\textit{Keywords ---}} #1}}}

\usepackage{natbib}
\usepackage{setspace}

\usepackage{thmtools}

\usepackage{authblk}

\usepackage{mathtools}
\mathtoolsset{showonlyrefs}

\usepackage[pdfstartview=XYZ,
bookmarks=true,
colorlinks=true,
linkcolor=blue,
urlcolor=blue,
citecolor=blue,
pdftex,
bookmarks=true,
linktocpage=true, 
hyperindex=true
]{hyperref}

\usepackage{orcidlink}

\author{Daniyar Ghani\,\orcidlink{0000-0001-8611-9966}}
\author{Nicholas A. Heard\,\orcidlink{0000-0002-8767-0810}}
\author{Francesco Sanna Passino\,\orcidlink{0000-0002-4571-6681}}

\affil{\small\it Department of Mathematics, Imperial College London\\
    \small\it 180 Queen's Gate, London SW7 2AZ, United Kingdom\\
    \small\it \href{mailto:daniyar.ghani18@imperial.ac.uk}{daniyar.ghani18@imperial.ac.uk}}

\date{}

\title{\huge\textbf{\titledoc}}

\allowdisplaybreaks

\numberwithin{equation}{section}

\begin{document}

\maketitle


\begin{abstract}
Models for categorical sequences typically assume exchangeable or first-order dependent sequence elements. These are common assumptions, for example, in models of computer malware traces and protein sequences. Although such simplifying assumptions lead to computational tractability, these models fail to capture long-range, complex dependence structures that may be harnessed for greater predictive power. 
To this end, a Bayesian modelling framework is proposed to parsimoniously capture rich dependence structures in categorical sequences, with memory efficiency suitable for real-time processing of data streams. Parsimonious Bayesian context trees are introduced as a form of variable-order Markov model with conjugate prior distributions. The novel framework requires fewer parameters than fixed-order Markov models by dropping redundant dependencies and clustering sequential contexts. Approximate inference on the context tree structure is performed via a computationally efficient model-based agglomerative clustering procedure. The proposed framework is tested on synthetic and real-world data examples, and it outperforms existing sequence models when fitted to real protein sequences and honeypot computer terminal sessions.
\end{abstract}


\section{Introduction}
\label{sec:intro}

Sequences of categorical data are ubiquitous in application areas such as language processing, bioinformatics and cyber-security. Statistical models for categorical sequences typically involve Markov assumptions \citep[see, for example,][]{Lewis2001, Maechler2004}, which yield easily interpretable and powerful methods for sequential prediction, changepoint detection and classification. Fixed order-$D$ Markov models are the natural extension of order-1 Markov chains, where the prediction of a new sequence element depends on the values of the previous $D$ elements, known as the \textit{context}. However, while fixed Markov models can theoretically learn rich dependence structures, they are limited by the exponential growth of parameter space. Given a state space, or vocabulary, of $V$ elements, a fixed order-$D$ Markov model requires $V^D(V-1)$ parameters.  The computational cost and storage requirements become excessively large as vocabulary size $V$ or Markov order $D$ grow. On the contrary, models using short contexts are computationally tractable but can have inferior predictive performance.  

To retain the ability to learn complex, long-range dependence structures while reducing the size of the parameter space, many parsimonious alternatives to fixed higher-order Markov models have been proposed in the literature. \cite{Rissanen1983} introduce variable-order Markov models (VOMM) for the purpose of data compression, where the Markov order depends on the context of the current element. Unlike fixed order-$D$ Markov models, the context lengths used for prediction in VOMMs can vary depending on the data. Inference algorithms for these models are explored in \cite{Begleiter2004}.

\cite{Bourguignon2004} extend variable-order Markov models in the form of a parsimonious context tree (PCT) structure, where nodes are fused together so that leaves may correspond to multiple contexts sharing the same predictive distributions. Developments in PCT learning algorithms are centred around dynamic programming \citep{Eggeling2013, Eggeling2019}, 
and while the efficiency of PCT learning has improved, the combinatorially large context trees required at initialisation limit the feasible vocabulary
to $V\approx 10$.  

Sparse Markov models \citep{Jaaskinen2014, Xiong2016, Bennett2022} and minimal Markov models \citep{Garcia2010, Garcia2017} are further generalisations of VOMMs with no hierarchical context tree structure; contexts of varying lengths are grouped into equivalence classes with the same predictive distributions. Such flexible modelling can uncover interesting dependence structures, yet inference remains challenging.  

Bayesian context trees \citep[BCT,][]{Kontoyiannis2022, Papageorgiou2022} are a
recent area of research focusing on exact Bayesian inference for VOMMs. BCT inference algorithms can efficiently recover long-range dependencies, and enable computation of quantities such as the prior predictive likelihood and model posterior probabilities, useful for downstream analysis. However, BCTs are still limited to small vocabularies. 

In this article, parsimonious Bayesian context trees (PBCT) are introduced as a new class of variable-order Markov model. The main differentiator of the proposed model with existing Bayesian context trees is the clustering of contexts, as the vocabulary is repeatedly partitioned in the tree to yield a variable-order Markov model over clusters. The advantages of context clustering are two-fold: a significant reduction in dimensionality and the ability to borrow statistical strength across similar contexts. To generate the structure of a PBCT, a natural recursive partitioning procedure is developed using the Chinese restaurant process \citep[CRP,][]{Aldous1985}. Dirichlet priors are placed on the unknown predictive distributions associated with each leaf of the tree to allow simple evaluation of the marginal likelihood of a sequence $\bm{x}$ given a context tree structure.

A novel algorithm is developed for efficient approximate inference of parsimonious Bayesian context trees, using model-based agglomerative clustering. The method proposed for inference enables the analysis of data with larger vocabulary sizes compared to existing inference schemes for similar context tree models. Results are obtained using synthetic sequences as well as real-world examples in cyber-security and bioinformatics, demonstrating the ability of the PBCT model to perfectly recover simulated model structures and outperform existing models for sequential prediction.

In Section~\ref{sec:context-trees}, the parsimonious Bayesian context tree model is described, including the assumed generative process. Section~\ref{sec:inference} outlines the model-based clustering inference scheme for learning PBCTs. Results of a simulation study are presented in Section~\ref{sec:sim-study}, and real data examples are investigated in Section~\ref{sec:apply-real}. 

\section{Parsimonious Bayesian context trees}
\label{sec:context-trees}

\subsection{Markov models}
\label{sec:markov-context}

Consider a discrete vocabulary $\mathcal{V}=\{1,2,\dots,V\}$, $V\geq 1$, and a potentially infinite sequence $x_1, x_2, \dots$, where each $x_i \in \mathcal{V}$. Suppose the first $N$ sequence elements have been observed as $\bm{x}=(x_1,\dots,x_N) \in \mathcal{V}^N$. For $D\geq 0$, an order-$D$ Markov model predicts the next element of the sequence $x_{N+1}$ conditional on the previous $D$ elements:
\begin{equation}
p(x_{N+1}\,|\,x_N,\dots,x_1) = 
\begin{cases}
  p(x_{N+1}), & \text{if } D=0, \\
  p(x_{N+1}\,|\,x_{N},\dots,x_{N-D+1}), & \text{if } D\geq 1.
\end{cases}\label{eq:pxN1}
\end{equation}
Equivalently, the next element depends on its length-$D$ context $(x_{N},\dots, x_{N-D+1})\in \mathcal{V}^D$.  

Markov models can be represented naturally using context trees, whose formal definition is given in Section~\ref{sec:pbct}. The root node of the tree represents the unobserved next element $x_{N+1}$ of a sequence, and the non-root nodes represent possible values of observed elements. The depth of a node corresponds to context length. In context tree representations of standard Markov models, nodes contain single elements of the vocabulary $\mathcal{V}$. In this article, parsimonious context trees are introduced as representations of parsimonious Markov models \citep{Bourguignon2004}, where nodes contain subsets of elements of $\mathcal{V}$. Each path from the root to a leaf node represents a collection of contexts which share the same predictive distribution for the next sequential element.  

\subsection{Context trees}
\label{sec:pbct}

A tree $\bm{\mathcal{T}}=(\bm{\mathcal{C}},\,\bm{\mathcal{E}})$ is an undirected graph with no cycles, where $\bm{\mathcal{C}}$ denotes the nodes of the graph and $\bm{\mathcal{E}}\subset \bm{\mathcal{C}}\times\bm{\mathcal{C}}$ is a set of edges. In predictive models for sequences over a vocabulary $\mathcal{V}$, the nodes $\bm{\mathcal{C}}$ of a \textit{parsimonious context tree} $\bm{\mathcal{T}}$ each contain subsets of $\mathcal{V}$, and are indexed according to their position within the tree; specifically, each node is associated with an index in $\mathbb{N}^d$ for some $0\leq d \leq D$, whose length corresponds to its depth in the tree hierarchy, and $D\geq 0$ is the maximum depth of the tree.

Let $\mathbb{E}_D=\bigcup_{d=0}^D \mathbb{N}^d$ denote the set of all possible indices for nodes in $\bm{\mathcal{T}}$. Each node in $\bm{\mathcal{C}}$ will be uniquely referred to as $C_{\bm{e}}$, for an index $\bm{e} \in \mathbb{E}_D$. The root node $C\equiv C_{\emptyset}=\mathcal{V}$ is indexed by the empty tuple. For any node pair $(C_{\bm{e}},\,C_{\bm{e}'})\in\bm{\mathcal{E}}$, where $\bm{e},\bm{e}' \in \mathbb{E}_D$, the node $C_{\bm{e}'}$ is said to be a \textit{child} of $C_{\bm{e}}$, and  $\bm{e}'$ must be the concatenation $(\bm{e},k)$ for some $k\in\mathbb{N}$. Therefore, if $C_{\bm{e}}$ has $K\geq 1$ children, these are indexed by sequences in $\{\bm{e}\} \times \{1,\dots,K\}$, written $\mathcal{C}_{\bm{e}} = \text{children}(C_{\bm{e}})=  \{C_{\bm{e}1}, C_{\bm{e}2}, \dots, C_{\bm{e}K}\}$. Crucially, every set $\mathcal{C}_{\bm{e}}$ of child nodes must form a partition of $\mathcal{V}$, so that $\bigcup_{k=1}^K C_{\bm{e}k} = \mathcal{V}$ and $C_{\bm{e}k}\cap C_{\bm{e}k'}=\emptyset$ for any $k\neq k'$. A \textit{leaf} node has no children, and each leaf node is associated with a unique conditional probability distribution \eqref{eq:pxN1} for the next element in a sequence, given a context matching the corresponding path through the tree.

\begin{figure}[t]
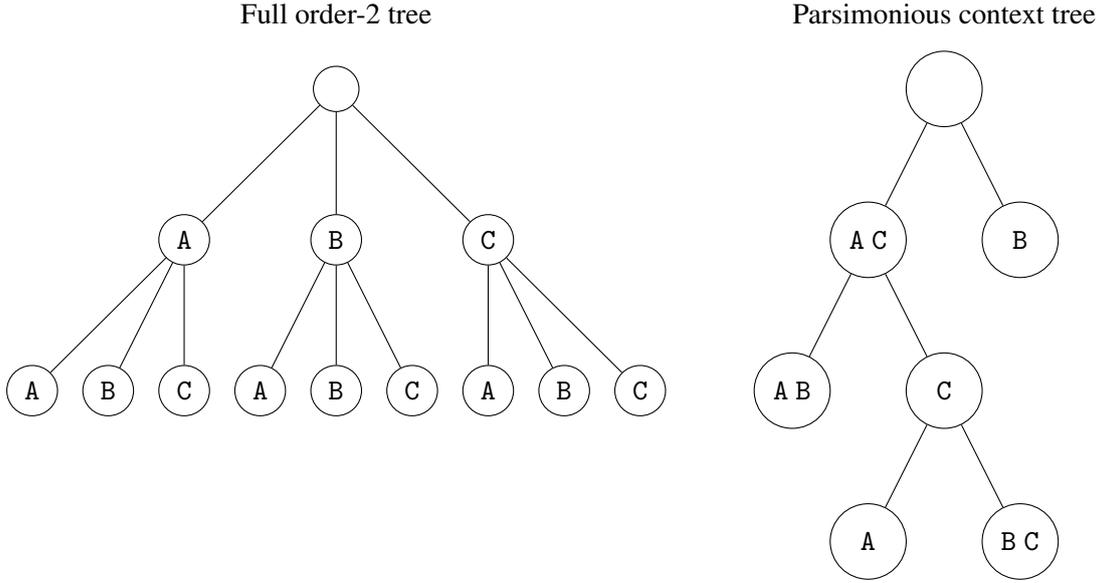

  \centering
  \ctikzfig{tree_comparison}
  \caption{\label{fig:tree-comparison} Comparison of tree structures for fixed order-2 Markov and parsimonious context tree models over the vocabulary $\mathcal{V}=\{\texttt{A},\texttt{B},\texttt{C}\}$.}
\end{figure}

Figure~\ref{fig:tree-comparison} compares the tree structures representing a fixed order-2 Markov model ($D=2$) and a parsimonious context tree with $D=3$. To understand how the tree structures determine sequential prediction, consider a sequence $\bm{x}\in \mathcal{V}^N$ over a 3-term vocabulary $\mathcal{V}=\{\texttt{A},\texttt{B},\texttt{C}\}$. Suppose the last three elements of the sequence $\bm{x}$ are observed as $(x_{N},x_{N-1},x_{N-2})=(\texttt{A},\texttt{C},\texttt{B})$. In the order-2 model, the distribution of the next element is given by $p(x_{N+1}\,|\,\bm{x}) = p(x_{N+1}\,|\,x_{N}=\texttt{A},\, x_{N-1}=\texttt{C})$, corresponding to the path: \textit{root} $\to$ \texttt{A} $\to$ \texttt{C}. Instead, in the parsimonious model in Figure~\ref{fig:tree-comparison}, the predictive distribution corresponds to the path: \textit{root} $\to$ \{\texttt{A},\,\texttt{C}\} $\to$ \{\texttt{C}\} $\to$ \{\texttt{B},\,\texttt{C}\}, and is given by $p(x_{N+1}\,|\,x_{N}\in\{\texttt{A},\texttt{C}\},\, x_{N-1}\in\{\texttt{C}\},\, x_{N-2} \in \{\texttt{B}, \texttt{C}\})$. 
The comparison illustrates the desired dimensionality reduction from fixed to parsimonious Markov models: a fixed order-3 model implies a tree with 27 leaves, each associated with a different predictive distribution, whereas the parsimonious model in Figure~\ref{fig:tree-comparison} defines only 4 predictive distributions. 

\subsection{Model specification}
\subsubsection{Conditional distributions}
\label{sec:cond-dists}
Given the parsimonious context tree $\bm{\mathcal{T}}$ with nodes $\bm{\mathcal{C}}$, let $E\subseteq \mathbb{E}_D$ denote the set of leaf node indices. Assign to each leaf node $C_{\bm{e}}$, $\bm{e}\in E$, a probability mass function $\bm{\phi}_{\bm{e}}$ over the vocabulary $\mathcal{V}$ with Dirichlet prior distributions: 
\begin{equation}
  \bm{\phi}_{\bm{e}} \sim \text{Dirichlet}(\bm{\eta}), \label{eq:prior}
\end{equation}
where $\bm{\eta} \in \mathbb{R}^{V}_{+}$. The tree $\bm{\mathcal{T}}$ and the implied conditional distributions $\{\bm{\phi}_{\bm{e}}\}_{\bm{e}\in E}$ define a \textit{parsimonious Bayesian context tree (PBCT)} model.  

\begin{figure}[t]
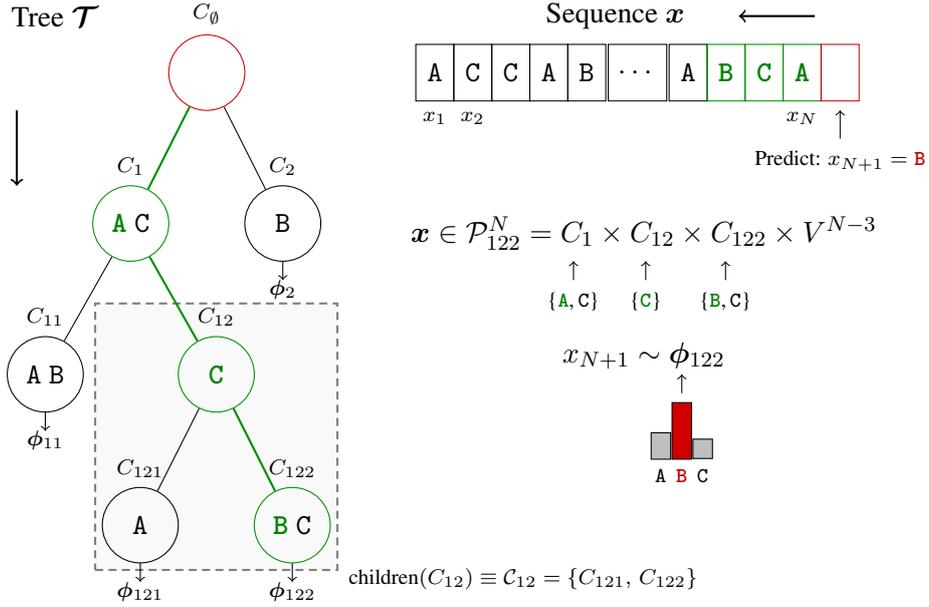

  \centering
  \ctikzfig{tree_example}
  \caption{\label{fig:tree-example} Example workings of the proposed Markov model for predicting the next element given a length-$N$ sequence $\bm{x}$ and tree $\bm{\mathcal{T}}$ with 4 leaves. The vocabulary is $\mathcal{V}=\{\texttt{A},\texttt{B},\texttt{C}\}$. Read the sequence from right to left and traverse the tree starting from the root (in red). The sequence corresponds to a path from root to a leaf via nodes $C_1$, $C_{12}$ and $C_{122}$. This path admits the predictive distribution $\bm{\phi}_{122}$ from which the next element $x_{N+1}$ is sampled.}
\end{figure}

For each leaf index $\bm{e}=(\epsilon_1, \epsilon_2, \dots, \epsilon_d)\in E$, define
\begin{equation}
  \mathcal{P}_{\bm{e}}^N = \{\bm{x} \in \mathcal{V}^N : x_N \in C_{\epsilon_1},\, x_{N-1} \in C_{\epsilon_1 \epsilon_2}, \dots, \,x_{N-d+1} \in C_{\bm{e}}\} \label{eq:partition}
\end{equation}
to be the set of $\bm{x}\in \mathcal{V}^N$ such that the length-$d$ context $(x_N, x_{N-1}, \dots, x_{N-d+1})$ matches the path from the root of $\bm{\mathcal{T}}$ to the leaf node indexed by $\bm{e}$. 
Suppose the leaf corresponding to sequence $\bm{x}$ has index $\bm{e}^{\bm{x}}\in E$, such that $\bm{x}\in \mathcal{P}_{\bm{e}^{\bm{x}}}^N$. Then the next element $x_{N+1}$ is sampled from the conditional distribution $\bm{\phi}_{\bm{e}^{\bm{x}}}$:
\begin{equation}
  x_{N+1}\,|\,\bm{x}, \bm{\mathcal{T}}, \{\bm{\phi}_{\bm{e}}\}_{\bm{e}\in E} \sim \text{Categorical} (\bm{\phi}_{\bm{e}^{\bm{x}}}), \label{eq:sim-next-element}
\end{equation}
yielding the Markov property $p(x_{N+1}\, | \, \bm{x}) = \bm{\phi}_{\bm{e}^{\bm{x}}}$ for sequential prediction. Figure~\ref{fig:tree-example} gives an illustrative example of next-element prediction given a PBCT $\bm{\mathcal{T}}$ and sequence $\bm{x}$.

\subsubsection{Marginal likelihood}
Consider a PBCT $\bm{\mathcal{T}}$ with leaf indices $E$. For $\bm{x} \in \mathcal{V}^N$, define $\bm{X}_{\bm{e}}=(X_{\bm{e}, 1}, \dots, X_{\bm{e}, V})$, where
\begin{equation}
  \label{eq:element-counts}
  X_{\bm{e}, v} = \sum_{n=1}^{N} \mathds{1}_{\mathcal{P}^{n-1}_{\bm{e}}}(\bm{x}_{n-1}) \mathds{1}_{\{v\}}(x_{n}).
\end{equation}
The count $X_{\bm{e}, v}$ is the number of times in $\bm{x}$ that $v\in \mathcal{V}$ follows a context indexed by $\bm{e}$, and the subsequence $\bm{x}_n=(x_1,\dots,x_n)$ denotes the first $n$ elements of $\bm{x}$. Due to the conjugacy of \eqref{eq:prior} with \eqref{eq:sim-next-element}, the marginal likelihood of $\bm{x}$ under $\bm{\mathcal{T}}$ is obtained from the Dirichlet--Categorical distribution:
\begin{equation}
  \label{eq:marginal-likelihood}
  p(\bm{x}\,|\,\bm{\mathcal{T}}) = \prod_{e\in E} \frac{B(\bm{X}_{\bm{e}} + \bm{\eta})}{B(\bm{\eta})},
\end{equation}
where $B(\bm{\alpha}) = \prod_i \Gamma(\alpha_i) / \Gamma(\sum_i \alpha_i)$ is the multivariate beta function.

In practice, for the simulation study in Section~\ref{sec:sim-study} and real-data applications in Section~\ref{sec:apply-real}, the counts $\bm{X}_{\bm{e}}$ are calculated ignoring the first $D$ elements of a sequence, where $D$ is the specified maximum tree depth. In this way, the sum of counts are computed over the values $D+1\leq n \leq $, avoiding any cases where the current subsequence $\bm{x}_n$ does not have enough sequential history to map to a leaf in the context tree.  

Another consideration for practical model inference is that discrete sequence data commonly occur as collections of many sequences over the same vocabulary. The computations of counts \eqref{eq:element-counts} and marginal likelihood \eqref{eq:marginal-likelihood} in this case are modified as follows: given a collection of observed sequences, the counts $X_{\bm{e}, v}$ for each sequence are calculated separately and aggregated, and the marginal likelihood
is computed using the aggregated counts.

\subsubsection{Generative process}
\label{sec:generation}
This section describes a method for generating a parsimonious Bayesian context tree graph. For each parent node in a PBCT, it is assumed that the child nodes correspond to a partition of the vocabulary $\mathcal{V}$, as described in Section~\ref{sec:pbct}. A natural approach to randomly partitioning a set of integers is the Chinese restaurant process \citep[CRP,][]{Aldous1985}, a common representation of the Dirichlet process \citep{Ferguson1973}. Suppose customers are to be seated at a restaurant with an infinite number of tables. Let $z_{m}\in \mathbb{N}$ denote the table allocation of customer $m$. The first customer sits at the first table, so $z_1=1$. For each successive customer $m\geq 2$, if $K$ tables have have already been occupied, the customer sits at an occupied table $k\in \{1,\dots,K\}$ or a new table $K+1$ with probabilities:
\begin{align*}
  \mathbb{P}(z_m=k) = \frac{m_k}{\alpha+m}, \quad \mathbb{P}(z_m=K+1) &= \frac{\alpha}{\alpha+m},
\end{align*}
where $\alpha>0$ controls the rate at which new tables are formed and $m_k$ denotes the number of customers seated at table $k$. By taking a finite realisation of the CRP after $V$ customers are seated, each occupied table will be associated with a subset of integers which together form an exchangeable partition of $\mathcal{V}=\{1,\dots,V\}$. The proposed generative process assumes a CRP prior distribution with parameter $\alpha$ over the set of partitions of $\mathcal{V}$, denoted $\text{CRP}_V(\alpha)$.
 
The following processes describe the generation of a PBCT $\bm{\mathcal{T}}=(\bm{\mathcal{C}},\,\bm{\mathcal{E}})$. Let $\mathcal{A}=\{A_1, \dots, A_K\}$ be a random partition of the vocabulary $\mathcal{V}$, where each $A_k \neq \emptyset$. For an infinite sequence of sets $(C_1,C_2,\dots)$, suppose $(C_1,C_2,\dots)\sim H_V(\alpha)$ if:
\begin{align}\label{eq:gen-H}
  \mathcal{A} \sim \text{CRP}_V(\alpha), \nonumber \quad
  C_k =
  \begin{cases}
    A_k,      & \text{if }  |\mathcal{A}|>1 \text{ and } k\leq |\mathcal{A}|, \\
    \emptyset,& \text{otherwise},
  \end{cases} \quad \text{for } k\in \mathbb{N}.
\end{align}
The process $H_V(\alpha)$ describes the construction of a sequence of sets starting with the components of a partition of $\mathcal{V}$ generated by the CRP, followed by a countably infinite number of empty sets. For $D\geq 1$, define $G_V(\alpha,D)$ as the process:
\begin{align}
    C \equiv C_{\emptyset} &= \mathcal{V}, \\
   (C_{\bm{e}1}, C_{\bm{e}2}, \ldots)
  &\sim
  \begin{cases}
    H_V(\alpha), & \text{if } C_{\bm{e}} \neq \emptyset,\\
    \delta_{(\emptyset,\emptyset,\ldots)},         & \text{otherwise},  
  \end{cases} \quad \text{for } \bm{e}\in \mathbb{E}_{D-1}, \\
  \bm{\mathcal{C}} &= \{C_{\bm{e}} \subseteq \mathcal{V} : C_{\bm{e}} \neq \emptyset,\, \bm{e}\in\mathbb{E}_D\}, \\
  \bm{\mathcal{E}} &= \{(C_{\bm{e}}, C_{\bm{e}k}) \in \bm{\mathcal{C}}\times\bm{\mathcal{C}} : \bm{e}\in\mathbb{E}_{D-1},\,k\in\mathbb{N}\}, \notag
\end{align}
where $\delta_{(\emptyset,\emptyset,\ldots)}$ denotes a Dirac measure placing probability one on the sequence $(\emptyset,\emptyset,\ldots)$. Then $\bm{\mathcal{T}}\sim G_V(\alpha,D)$ is the generative process of the tree $\bm{\mathcal{T}}$ of maximum depth $D$ with nodes $\bm{\mathcal{C}}$ and edges $\bm{\mathcal{E}}$. The children of each non-leaf node of the tree form a partition of $\mathcal{V}$.  

There are two ways to control the complexity of a generated tree: (i) vary the maximum depth of the tree, $D$, or (ii) vary the CRP rate parameter $\alpha$. Smaller values of $\alpha$ lead to partitions of $\mathcal{V}$ with fewer components, or clusters, constraining the complexity of the tree. Additionally, to limit tree size, $\alpha$ may be chosen to decay with depth, causing the expected number of generated child nodes to decrease with depth.

\section{Inference}
\label{sec:inference}
Section~\ref{sec:generation} described the assumed generative process of a parsimonious Bayesian context tree.
In this section, an approximate Bayesian inference scheme is described to learn PBCT structure given data. The introduced method, based on agglomerative clustering, is applied in Section~\ref{sec:sim-study} to infer synthetic tree structures, and in Section~\ref{sec:apply-real} to estimate the dependence structures in real protein sequences and command-line data.

\subsection{Agglomerative clustering}
\label{sec:agg-clust}
Agglomerative hierarchical clustering \citep{Duda1973} is a general method for partitioning a discrete set of elements using a similarity metric. The procedure starts with each element in a singleton cluster, and successively merges clusters according to their similarities until all elements are grouped together in the same cluster. A nested sequence of clusterings is created, and the optimal clustering is chosen to maximise an objective function. Bayesian model-based agglomerative schemes
\citep{Heller2005, Heard2006}, 
use marginal posterior probabilities to define similarities between clusters.

Model-based agglomerative clustering is applied to infer the parsimonious Bayesian context tree $\bm{\mathcal{T}}$ of maximum depth $D$ given a sequence $\bm{x}$. The clustering process is initialised at the root node and repeated recursively to obtain the optimal children of each node: each recursion terminates if the optimal child configuration is $\{\mathcal{V}\}$, or the current node is at the maximum depth $D$, at which point the current node is set as a leaf of the tree.  

Starting from the root node $C\equiv C_{\emptyset}$, let $\bm{e}\in \mathbb{E}_d$ be an index at tree depth $d\leq D$, and let $\mathcal{C}_{\bm{e}}=\{C_{\bm{e}1},\ldots,\,C_{\bm{e}K}\}$ denote the children of node $C_{\bm{e}}$ which form a partition of the vocabulary $\mathcal{V}$. Denote by $\bm{\mathcal{T}}_d$ the tree cut at the current depth $d$. The cluster similarity at each step is the multiplicative increase in probability obtained by merging $C_{\bm{e}i}$ and $C_{\bm{e}j}$:
\begin{equation}
  s_{i,j} = \frac{p( \bm{x}\, | \,\bm{\mathcal{T}}_d,\, \mathcal{C}_{\bm{e}}^{(i,j)}) \, p(\mathcal{C}_{\bm{e}}^{(i,j)})}{p(\bm{x}\,|\,\bm{\mathcal{T}}_d,\, \mathcal{C}_{\bm{e}})\,p(\mathcal{C}_{\bm{e}})},
\end{equation}
where $\mathcal{C}_{\bm{e}}^{(i,j)}$ denotes the cluster configuration in which clusters $C_{\bm{e}i}$ and $C_{\bm{e}j}$ are merged, and $p(\bm{x}\,|\, \bm{\mathcal{T}}_d,\, \mathcal{C}_{\bm{e}})$ is the marginal likelihood of $\bm{x}$ under the current tree $\bm{\mathcal{T}}_d$ with the proposed cluster configuration $\mathcal{C}_{\bm{e}}$ at depth $d+1$. Assuming the generative process in Section~\ref{sec:generation}, the prior probability $p(\mathcal{C}_{\bm{e}})$ is available from the Chinese restaurant process. The agglomerative clustering procedure for the children of each node $C_{\bm{e}}$ at depth $d$ starts with each element of $\mathcal{V}$ belonging to a singleton cluster, and is sequentially iterated until all elements are merged in the same group; the clustering structure $\mathcal{C}_{\bm{e}}$ which locally maximises the marginal posterior probability $\pi(\mathcal{C}_{\bm{e}}) = p(\bm{x}\, \vert\, \bm{\mathcal{T}}_d,\, \mathcal{C}_{\bm{e}})\, p(\mathcal{C}_{\bm{e}})$ is chosen as the optimal child configuration for the corresponding node in the tree. This scheme is denoted \emph{recursive agglomerative clustering} (RAC), and it is outlined in Algorithm~\ref{alg:agg-clust}.

\begin{algorithm}
\setstretch{1}
\caption{Recursive agglomerative clustering (RAC)}
\label{alg:agg-clust}
\begin{algorithmic}[1]
  \STATE \textbf{Initialise:} Set current node to root node: $C_{\bm{e}} \equiv C_{\emptyset} = \mathcal{V}$.
  \WHILE{current node $C_{\bm{e}}$ is not a leaf}
    \STATE Initialise clustering $\mathcal{C}_{\bm{e}} = \{C_{\bm{e}1},\ldots,C_{\bm{e}V}\},$ where each $C_{\bm{e}i}=\{i\}$.
    \STATE Set number of clusters $K=V$ and calculate $\pi_K = \pi(\mathcal{C}_{\bm{e}})$.
    \WHILE{$K>1$}
        \STATE Merge clusters $C_{\bm{e}i}$ and $C_{\bm{e}j}$ with maximal similarity $s_{i,j}$.
	\STATE Set $K=K-1$ and relabel clusters $i,j$ accordingly.
	\STATE Calculate $\pi_K = \pi(\mathcal{C}_{\bm{e}})$ for new clustering.
    \ENDWHILE
    \STATE Select the optimal clustering $\mathcal{C}_{\bm{e}}$ which maximises $\pi(\mathcal{C}_{\bm{e}})$.
    \IF{$\mathcal{C}_{\bm{e}}=\{\mathcal{V}\}$}
      \STATE Set current node $C_{\bm{e}}$ as a leaf.
    \ELSE
      \FOR{each child node $C_{\bm{e}'} \in \mathcal{C}_{\bm{e}}$}
        \STATE Set current node to child node: $C_{\bm{e}}=C_{\bm{e}'}$.
        \IF{$C_{\bm{e}}$ is at maximum depth}
        		\STATE Set current node $C_{\bm{e}}$ as a leaf.
	\ELSE
		\STATE Repeat the procedure from line 2 to get optimal children of current node $C_{\bm{e}}$.
	\ENDIF
      \ENDFOR
    \ENDIF
  \ENDWHILE
\end{algorithmic}
\end{algorithm}

The RAC algorithm provides a fast and scalable procedure for estimating the tree which generated an observed sequence. In Section ~\ref{sec:sim-study}, the RAC algorithm is shown to recover the true underlying context trees from simulated data in a variety of settings. An advantage of the RAC algorithm over existing methods is the ability to handle significantly larger vocabulary sizes than existing methods for context tree learning. For instance, the algorithms presented in \cite{Eggeling2019} require initialisation with extended trees containing all possible combinations of child node configurations up to the specified maximum tree depth. This limits feasible vocabulary size to $V\approx 10$, whereas RAC can learn PBCTs in reasonable computational time for data with $V\approx 100$, as shown in Section~\ref{sec:apply-real}.

\subsection{Model evaluation}
\label{sec:model-eval}
Suppose the PBCT $\bm{\mathcal{T}}$ is inferred from the training sequence $\bm{x}^{\text{train}}$. If $\bm{\mathcal{T}}$ has leaf indices $E$, the predictive marginal likelihood of an unseen length-$N$ test sequence $\bm{x}^{\text{test}} \in \mathcal{V}^N$ under the trained model is
\begin{align}
    p(\bm{x}^{\text{test}}\,|\,\bm{x}^{\text{train}},\,\bm{\mathcal{T}}\,) = \prod_{\bm{e}\in E} \frac{B(\bm{X}_{\bm{e}}^{\text{test}} + \bm{X}_{\bm{e}}^{\text{train}} + \bm{\eta})}{B(\bm{X}_{\bm{e}}^{\text{train}}+\bm{\eta})}
\end{align}
where $\bm{X}_{\bm{e}}^{\cdot}$ are the counts \eqref{eq:element-counts} calculated separately over the training and test sequences.
To evaluate the predictive performance of a trained model, define the \textit{marginal log-loss} $\ell(\bm{x}^{\text{test}})$ to be the mean negative predictive log-likelihood of the test sequence,
\begin{align}
\label{eq:log-loss}
    \ell(\bm{x}^{\text{test}} \, |\, \bm{x}^{\text{train}},\,\bm{\mathcal{T}}\,) = -\frac{1}{N} \log p(\bm{x}^{\text{test}} \,|\, \bm{x}^{\text{train}},\,\bm{\mathcal{T}}\,).
\end{align}
Lower marginal log-loss indicates superior predictive ability.

To obtain the marginal log-loss \eqref{eq:log-loss} for a given tree, the predictive distributions for each leaf are marginalised. For simulation studies, it is also possible to define a \textit{true log-loss} for a model for which the true conditional distributions, denoted $\{\bm{\phi}_{\bm{e}}^{*}\}$, are assumed known:
\begin{align}
\label{eq:true-log-loss}
    \ell(\bm{x}^{\text{test}} \,|\, \bm{x}^{\text{train}},\,\bm{\mathcal{T}},\, \{\bm{\phi}_{\bm{e}}^{*} \}_{\bm{e}\in E} ) 
    = -\frac{1}{N} \sum_{i=1}^{N} \log \phi^{*}_{\bm{e}^{\bm{x}_i},\,x_i}.
\end{align}
Estimates of the predictive distributions can be useful for efficient downstream analysis: for each leaf $\bm{e}\in E$, denote by $\hat{\bm{\phi}}_{\bm{e}}=(\hat{\phi}_{\bm{e},1}, \dots, \hat{\phi}_{\bm{e},V})$ the posterior mean estimates of the corresponding Dirichlet--Categorical distribution,
\begin{align}
 \hat{\phi}_{\bm{e}, v} = \frac{X_{\bm{e},v}^{\text{train}} + \eta_v}{\sum_{u=1}^V X_{\bm{e},u}^{\text{train}} + \eta_u}, \quad v\in \mathcal{V}.
\end{align}

A similarity metric based on the adjusted Rand index \citep[ARI,][]{Hubert1985} is defined to compare tree structures; for example, to compare a fitted tree with a known, simulated tree. For each clustering at depth $d\leq D$ in the fitted tree, the metric obtains the ARI of the most similar clustering at depth $d$ in the simulated tree; these maximal ARIs are then weighted by the frequencies of contexts in the training sequence and averaged. In this way, more importance is placed on the parts of the trees which are commonly represented in the data, and errors are penalised less for rare contexts.   

The tree similarity metric is defined as follows. Let $\bm{\mathcal{T}}_1$ and $\bm{\mathcal{T}}_2$ be trees of maximum depths $D_1$ and $D_2$, such that $D_1\geq D_2$. For $m\in\{1,2\}$, let $I_m \subseteq \mathbb{E}_{D_m}$ denote the indices of all nodes in $\bm{\mathcal{T}}_m$ and $E_m\subseteq I_m$ denote the indices of leaf nodes in $\bm{\mathcal{T}}_m$. For any node $\bm{e}_m \in I_m$ at depth $d_m$, let $\mathcal{C}_{\bm{e}_m}$ denote the partition of $\mathcal{V}$ comprising the children of node $C_{\bm{e}_m}$. If $\bm{e}_m \in E_m$ is a leaf, then let $\mathcal{C}_{\bm{e}_m}=\{\mathcal{V}\}$. Define the weight $w_{\bm{e}_1}$ as the proportion of contexts in $\bm{x}$ associated with node $C_{\bm{e}_1}$ over all nodes at the same depth in tree $\bm{\mathcal{T}}_1$, and let $\text{ARI}_{\bm{e}_1, \bm{e}_2}$ denote the adjusted Rand index between clusterings $\mathcal{C}_{\bm{e}_1}$ in $\bm{\mathcal{T}}_1$ and $\mathcal{C}_{\bm{e}_2}$ in $\bm{\mathcal{T}}_2$. Then, for $0\leq d \leq D_1-1$, define the \textit{tree similarity} at depth $d+1$ as
\begin{equation}
\label{eq:ari-metric}
\sum_{\bm{e}_1 \in \mathbb{N}^{d}\, \cap \, I_1}
		    w_{\bm{e}_1} 
		    \max_{\bm{e}_2 \in \mathbb{N}^{d} \, \cap\, I_2}
		    \left\{ \text{ARI}_{\bm{e}_1,\bm{e}_2}\right\}.
\end{equation}


\section{Simulation study}
\label{sec:sim-study}
A simulation study enables assessment of the performance of the inference procedure for learning a parsimonious Bayesian context tree (PBCT). This section explores two experiments: (i) a hyperparameter study, and (ii) a model comparison as training lengths vary.

\subsection{Setup}
In both experiments, PBCT models are randomly generated over the vocabulary $\mathcal{V}$ as described in Section~\ref{sec:generation}, and sequences are simulated using these trees. Each simulated sequence is split into a training sequence $\bm{x}^{\text{train}}$ and test sequence $\bm{x}^{\text{test}}$. Given a simulated sequence, the task is to recover the true generated tree. Trees are learned from the training sequence via the recursive agglomerative clustering (RAC) procedure described in Section~\ref{sec:agg-clust}, Algorithm~\ref{alg:agg-clust}. The evaluation metrics described in Section~\ref{sec:model-eval} -- marginal log-loss \eqref{eq:log-loss}, true log-loss \eqref{eq:true-log-loss} and tree similarity \eqref{eq:ari-metric} -- are used to evaluate model performance. 

The first experiment investigates the sensitivity of RAC to the simulation parameter $\bm{\eta}\in \mathbb{R}^V_{+}$. Additionally, the simulated predictive distributions are modified to capture the influence of a background level of stochastic variation in a sequence. Let $\lambda \in [0,1]$ be a parameter to control the strength of the stochastic variation. Then, the conditional distribution $\bm{\phi}_{\bm{e}}^{\text{true}}$ for each leaf $\bm{e}$ of a simulated tree is considered to be a mixture of the $\text{Dirichlet}(\bm{\eta})$ distribution and the uniform distribution over $\mathcal{V}$: the simulated distribution is given by $\bm{\phi}_{\bm{e}}^{\text{true}} = (1-\lambda) \bm{\phi}_{\bm{e}} + \lambda \mathbf{1}_{V}/V, \,\,  \bm{\phi}_{\bm{e}} \sim \text{Dirichlet}(\bm{\eta})$, where $\mathbf{1}_{V}$ is a length-$V$ vector of ones. In this ``spike-and-slab'' representation of $\bm{\phi}_{\bm{e}}^{\text{true}}$, $\lambda=0$ yields the standard Dirichlet distribution, and $\lambda=1$ yields the uniform distribution. 

Synthetic parsimonious Bayesian context trees are generated over a vocabulary of size $V=10$, and each partition of the vocabulary is drawn from $\text{CRP}_{V}(\alpha)$ with rate parameter $\alpha=1$. The maximum depth of the PBCTs is specified as $D=3$. Training sequences of length $N=10,000$ and test sequences of length $N=1,000$ are simulated from the generated trees. The Dirichlet hyperparameters are specified as $\bm{\eta}=\eta \mathbf{1}_V$, and the simulation parameters $\eta$ and $\lambda$ are varied. Results are obtained for 15 different model simulations. Model performance is evaluated by computing the ARI-based similarity metric \eqref{eq:ari-metric} over the training sequences to compare tree structures, and marginal log-losses \eqref{eq:log-loss} are computed over the test sequences to evaluate predictive performance.

In the second experiment, fixed and variable-order Markov models are considered for comparison with the PBCT. Let $D$ be the maximum tree depth, so PBCT-$D$ denotes a parsimonious Bayesian context tree of maximum depth $D$. A fixed order-$d$ Bayesian Markov model (FBM-$d$), $d\in \{1,\dots,D\}$, is defined such that Dirichlet priors are placed on the conditional distributions for each order-$d$ context. Similarly, a variable-order Bayesian Markov model with maximum order $D$ \citep[VBM-$D$; see, for example,][]{Dimitrakakis2010} is defined with Dirichlet conditional distributions for each context. VBM structure is inferred using a modified version of RAC, Algorithm~\ref{alg:agg-clust}: at each vocabulary partitioning step, only two outcomes are compared: (i) all elements in the same cluster, or (ii) $V$ singleton clusters.

Predictive performance is investigated as the length of training sequence varies. The hyperparameter pair $\bm{\eta}=\mathbf{1}_V$ and $\lambda=0$ are used for sequential prediction so that the predictive distributions associated with each leaf are standard Dirichlet distributions. The synthetic models are generated with vocabulary size $V=10$, CRP parameter $\alpha=1$ and maximum depth $D=3$. Predictive log-losses are averaged over 15 model simulations. 

\begin{figure}[t!]
\begin{center}
\includegraphics[width=0.9\textwidth]{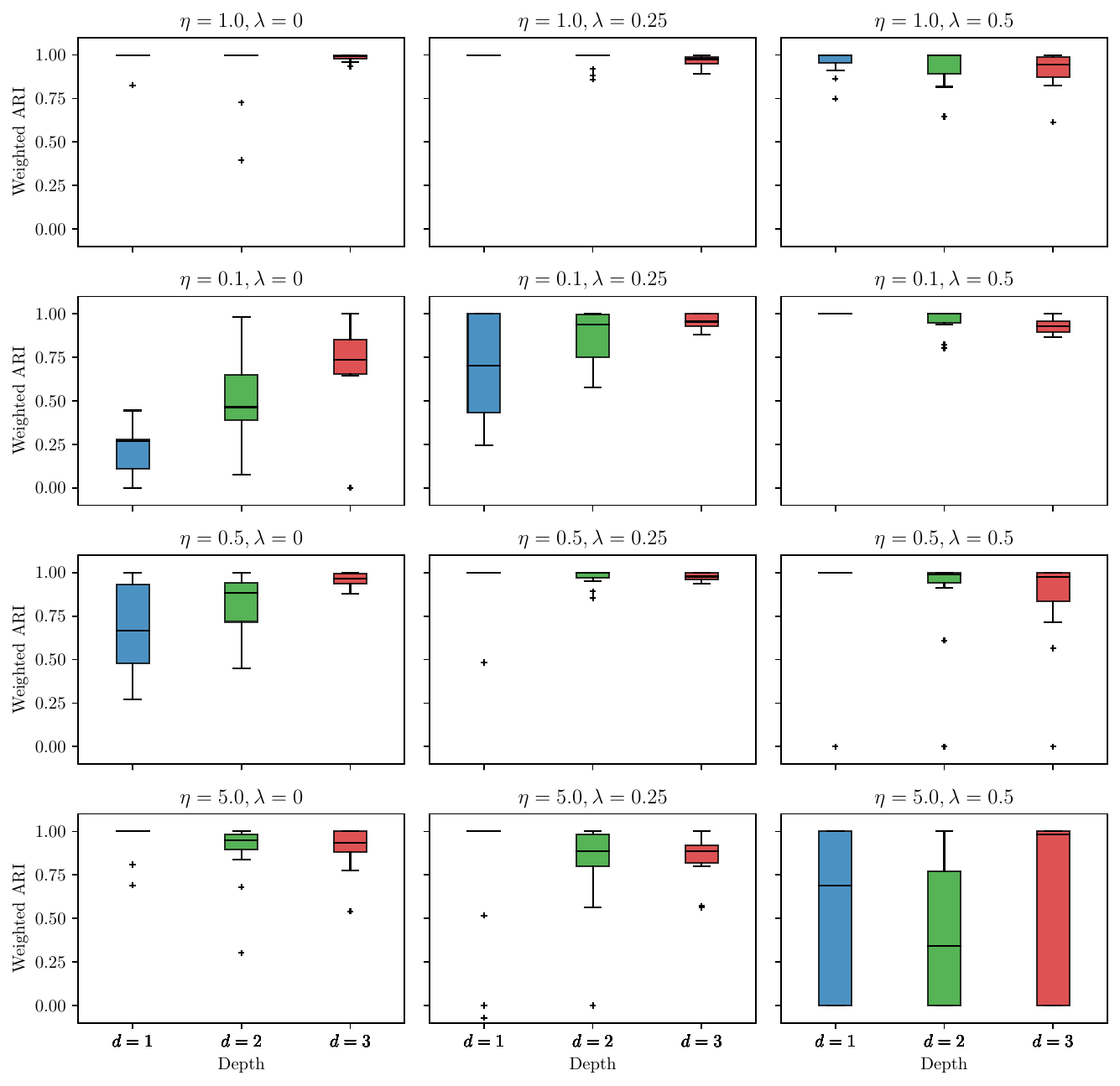}
\end{center}
\caption{Boxplots to compare structures of fitted and simulated trees. Synthetic trees are generated with vocabulary size $V=10$, CRP parameter $\alpha=1$ and maximum depth $D=3$. $N=10,000$ training samples are simulated using the trees for varying distributional hyperparameters $\eta$ and $\lambda$. Boxplots of weighted ARI for each depth $d=1,2,3$ over 15 different model simulations.
\label{fig:ari-boxplots}}
\end{figure}

\subsection{Results}
Figure~\ref{fig:ari-boxplots} illustrates boxplots of tree similarities for different simulation parameter configurations, and Table \ref{tab:sim-study-loss-table} contains the average log-losses of the simulated and fitted trees for the first experiment. It is found that the RAC algorithm for learning PBCTs generally performs well, and can perfectly recover synthetic context tree structures for several parameter configurations. The best and most consistent performance is achieved by the hyperparameter pair $\eta=1, \, \lambda=0$. It can be seen that the presence of random noise can lead to improved performance for small values of $\eta$. Small $\eta$ values lead to highly sparse Dirichlet distributions, leading to issues with model identifiability. Adding noise smooths the conditional distributions to ensure greater variation of contexts in the simulated sequence, which helps RAC recover the simulated tree. On the other hand, as $\eta$ becomes large, the predictive distributions each become close to uniform, again leading to identifiability issues.

\begin{table}[t]
  \centering
  \caption{Average log-losses for simulated and fitted PBCT models for varying distributional hyperparameter pairs $\eta$ and $\lambda$ repeated over 15 models. Standard deviations in parentheses. Synthetic trees generated with vocabulary size $V=10$, CRP parameter $\alpha=1$ and maximum depth $D=3$. Sequences of length $N=10,000$ are simulated from each tree. Best performance is given by the hyperparameters $\eta=1$, $\lambda=0$.}
    \begin{tabular}{llccc}
     \toprule
                $\eta$   & $\lambda$ & Simulated & Fitted & Difference \\ 
    \midrule
    1.0      & 0.0       & 1.9297 (0.0674) & 1.9407 (0.0667) & \textbf{0.0017 (0.0023)} \\
    1.0      & 0.25      & 2.1158 (0.0501) & 2.1269 (0.0538) & 0.0093 (0.0134) \\
    1.0      & 0.5       & 2.2177 (0.0168) & 2.2290 (0.0211) & 0.0090 (0.0099) \\
    \midrule
    0.1      & 0.0       & 0.9045 (0.1314) & 1.0249 (0.2631) & 0.1236 (0.2117) \\
    0.1      & 0.25      & 1.5552 (0.1408) & 1.5921 (0.1988) & 0.0477 (0.1123) \\
    0.1      & 0.5       & 1.9397 (0.0337) & 1.9436 (0.0372) & \textbf{0.0042 (0.0054)} \\
    \midrule
    0.5      & 0.0       & 1.6800 (0.0888) & 1.6979 (0.0993) & 0.0214 (0.0252) \\
    0.5      & 0.25      & 1.9587 (0.0693) & 1.9656 (0.0669) & \textbf{0.0088 (0.0076)} \\
    0.5      & 0.5       & 2.1503 (0.0336) & 2.1681 (0.0456) & 0.0231 (0.0309) \\
    \midrule
    5.0      & 0.0       & 2.2165 (0.0262) & 2.2271 (0.0257) & 0.0110 (0.0109) \\
    5.0      & 0.25      & 2.2581 (0.0160) & 2.2716 (0.0161) & \textbf{0.0089 (0.0099)} \\
    5.0      & 0.5       & 2.2871 (0.0054) & 2.2959 (0.0073) & 0.0094 (0.0067) \\
    \bottomrule
      \end{tabular}
  \label{tab:sim-study-loss-table}
\end{table}

\begin{figure}[t!]
    \centering
    \begin{subfigure}[b]{0.49\textwidth}
    	\centering
    	\includegraphics[width=\textwidth]{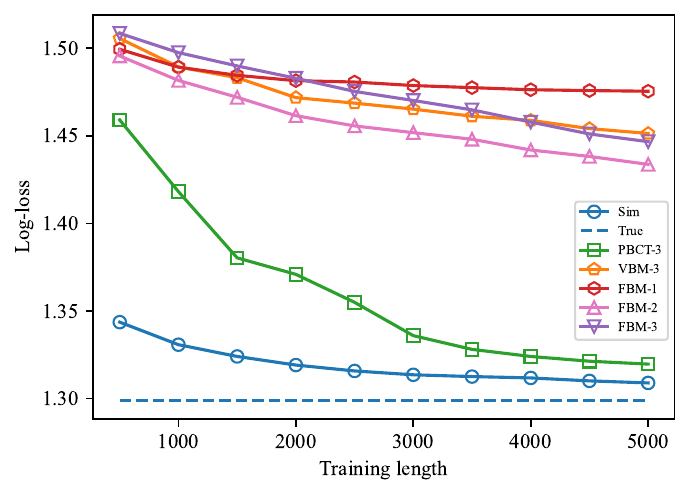}
	\caption{\label{fig:loss-train}}
    \end{subfigure}
    \hfill 
    \begin{subfigure}[b]{0.49\textwidth}
    	\centering
    	\includegraphics[width=\textwidth]{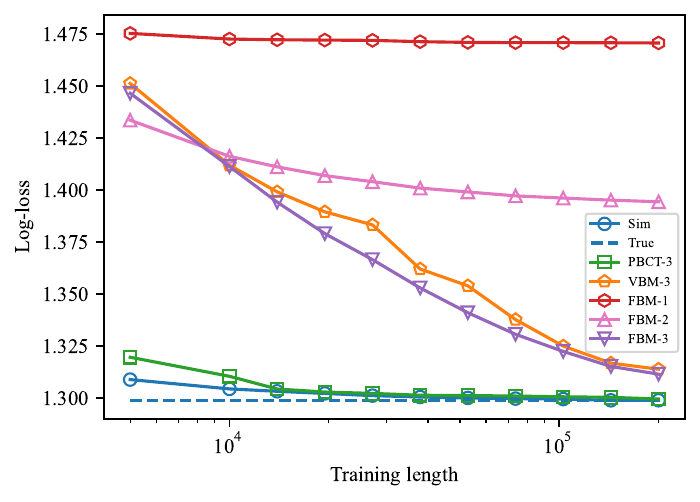}
	\caption{\label{fig:loss-train-long}}
    \end{subfigure} \\
    \vfill
    \begin{subfigure}[b]{0.49\textwidth}
    	\centering
    	\includegraphics[width=\textwidth]{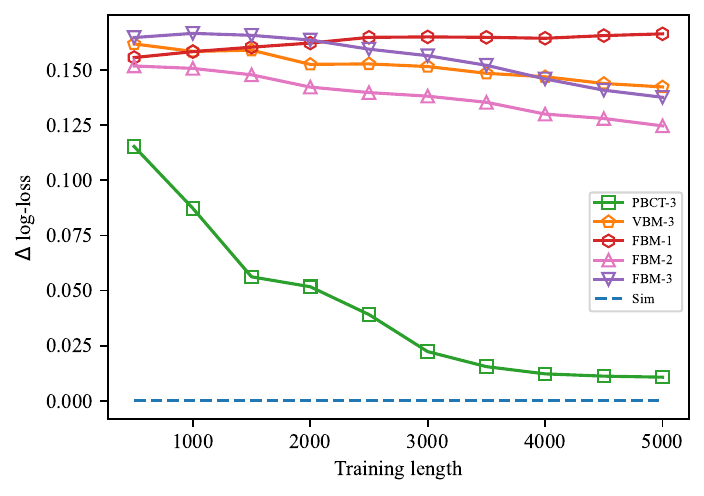}
	\caption{\label{fig:delta-loss-train}}
    \end{subfigure} 
    \hfill 
    \begin{subfigure}[b]{0.49\textwidth}
    	\centering
    	\includegraphics[width=\textwidth]{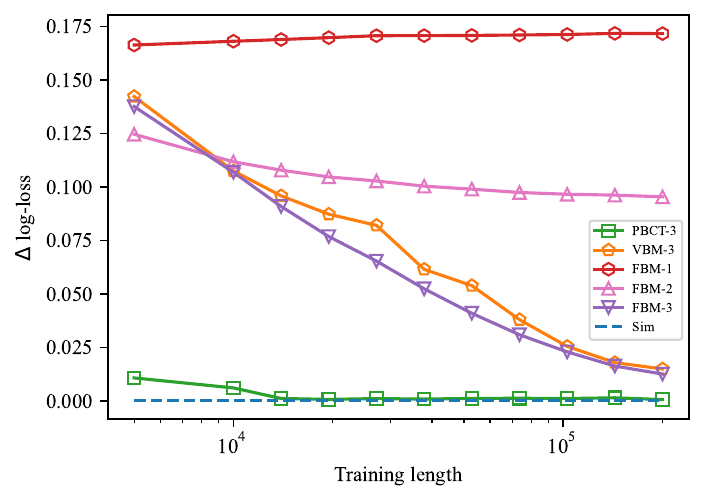}
	\caption{\label{fig:delta-loss-train-long}}
    \end{subfigure}
    \caption{Average fitted log-losses and difference in log-losses between fitted and simulated models averaged over 15 simulations, for varying training lengths and a fixed test length of 1000 elements. Synthetic trees are generated with vocabulary size $V=10$, CRP parameter $\alpha=1$ and maximum depth $D=3$. (\ref{fig:loss-train}) and (\ref{fig:delta-loss-train}) show PBCT outperforms the other Markov models over all training lengths. (\ref{fig:loss-train-long}) and (\ref{fig:delta-loss-train-long}) show consistent recovery of simulated PBCT structure for longer training lengths.}
    \label{fig:loss-train-all}
\end{figure}

Figure~\ref{fig:loss-train-all} displays average log-losses for a selection of fitted models, as well as the difference in log-losses with the simulated models, as training length varies. Figures~\ref{fig:loss-train} and \ref{fig:delta-loss-train} show the log-losses for training lengths up to $N=5,000$, where the PBCT outperforms fixed and variable-order Markov models for all training lengths. Additionally, Figures~\ref{fig:loss-train-long} and \ref{fig:delta-loss-train-long} illustrate continuations of the same experimental simulations for longer training lengths, plotted on a log-scale, from $N=5,000$ up to $N=200,000$. The PBCT model consistently recovers the simulated trees given longer training sequences. The fixed and variable-order Markov models, FBM-3 and VBM-3, continue to improve with longer training lengths but do not reach the same performance as PBCT-3.

\section{Application to real data}
\label{sec:apply-real}

The parsimonious Bayesian context tree model is applied to two real-world examples of categorical sequence data: (i) Imperial honeypot terminal sessions and (ii) UniProt protein sequences. Comparisons are made with fixed and variable-order Markov models.

\begin{table}[b!]
    \centering
    \caption{Sequence statistics for real-world datasets.}
    \begin{tabular}{lcccc}
        \toprule
        Dataset & Mean length & Total training length & Total test length \\
        \midrule
        Honeypot & 52 & 46225 & 5402 \\
        UniProt & 69 & 31167 & 3370\\
        \bottomrule
    \end{tabular}
    \label{tab:eda-stats}
\end{table}

\begin{table}[t]
    \centering
     \caption{Predictive performance of each model evaluated on the honeypot and UniProt datasets. Best log-loss performance is achieved by PBCT for both datasets.}
    \begin{tabular}{p{1.7cm}@{}cp{1.7cm}@{}cp{1.7cm}}
        \toprule
        \multirow{2}{*}{Model} & \multicolumn{2}{c}{Honeypot} & \multicolumn{2}{c}{UniProt} \\
        \cmidrule(lr){2-3} \cmidrule(lr){4-5}
        & $L$ & log-loss & $L$ & log-loss \\
        \midrule
        FBM-0 & 1 & 2.72327 & 1 & 2.83083 \\
        FBM-1 & 93 & 1.04483 & 21 & 2.57862 \\
        FBM-2 & 8,649 & 0.72451 & 441 & 1.89768 \\
        FBM-3 & 804,357 & - & 9,261 & 1.48856 \\
        FBM-4 & - & - & 194,481 & 1.52664 \\
        VBM & 93 & 1.04483 & 5,641 & 1.48510 \\
        BCT\protect\footnotemark & 1,312   & 0.70253 & 3,820   & 1.48539 \\
        PBCT & 654 & \textbf{0.69012} & 1,076 & \textbf{1.47870} \\
        \bottomrule
    \end{tabular}
    \label{tab:combined-results}
\end{table}

\subsection{Honeypot terminal sessions}
\label{sec:honeypot}
In this example, the data are sessions of command-line sequences collected on a honeypot at Imperial College London. A \textit{honeypot} is a type of computer network host designed to observe the behaviour of cyber attackers when granted access to the network. The Imperial honeypot data are in the form of sessions, where each session is formed of a sequence of Unix terminal commands executed by an attacker. A collection of 1000 sessions, recorded between May 2021 and January 2022, are considered for model training and evaluation. 90\% of the sessions are used for training and the remaining 10\% are used to evaluate predictive performance. Table \ref{tab:eda-stats} details the average session lengths and total number of commands in the data.  
\footnotetext{BCTs were fitted to concatenations of sequences; the software did not accept multiple sequence inputs.}
From the honeypot sessions, each command is considered a word in the vocabulary $\mathcal{V}$. After preprocessing commands into common categories and key Unix commands, the vocabulary has $V=93$ elements. The PBCT model is trained using CRP rate parameter $\alpha=1$ and Dirichlet hyperparameter $\bm{\eta}=\mathbf{1}_V$. Fixed Bayesian Markov models of orders $D\in\{0,1,2,3\}$ are defined (FBM-$D$, where FBM-0 is a bag-of-words model) and the variable-order Bayesian Markov (VBM) model is described in Section~\ref{sec:sim-study}. For comparison, a Bayesian context tree (BCT) is fitted using the authors' publicly available software \citep{Kontoyiannis2022}. For each model, the conditional distributions $\bm{\phi}_{\bm{e}}$ used for prediction are Dirichlet with parameter $\bm{\eta}=\mathbf{1}_V$. A maximum depth of $D=3$ is specified before training the variable-order models PBCT, VBM and BCT. 

Table \ref{tab:combined-results} outlines the means and standard deviations of log-losses computed on the test sequences for each model. The number of conditional distributions defined by each model is denoted by $L\in\mathbb{N}$. The best predictive performance is achieved by the PBCT. The fitted PBCT has $L=654$ and a maximum depth of 2, exhibiting a significant reduction in dimensionality while achieving better predictive performance than the fixed order-2 Markov model (FBM-2), which has $L=8,649$. The fitted BCT, also of maximum depth $D=2$, has $L=1,132$ and is slightly outperformed by the PBCT in terms of predictive log-loss. The log-loss for FBM-3 is omitted due to excessive computational cost ($L=804,357$).  

\begin{figure}[h]
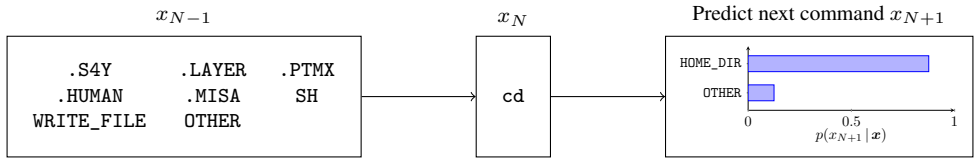

  \centering
  \ctikzfig{honeypot-interpretations}
   \caption{Example of learned command contexts and prediction using the PBCT fitted to honeypot sessions.}
   \label{fig:honeypot-interpret} 
\end{figure}

Figure~\ref{fig:honeypot-interpret} illustrates an example of a learned order-2 context in the PBCT and the two largest predictive probabilities, calculated as posterior means. The example demonstrates the ability to predict the behaviour of a network intruder: an attacker first attempts installation of a MIRAI variant \citep[such as \texttt{.PTMX} or \texttt{.MISA}; see][for further discussion]{SannaPassino2023}, then changes directory using the \texttt{cd} command. Following this context, the PBCT predicts the common behaviour of navigating back to the home directory.  

\subsection{Protein sequences}
\label{sec:uniprot}
The UniProt knowledgebase \citep{UniProt2016} is a large collection of protein sequences. Each entry is a sequence with elements from a vocabulary of the 20 standard amino acids. There is an extra element added to the vocabulary in the case of a missing or erroneous amino acid, so the final vocabulary has size $V=21$. A random sample of sequences of lengths between 10 and 100 is considered for computational feasibility. Table \ref{tab:eda-stats} details the average and total protein sequence lengths used for analysis.

The PBCT model is fitted to the protein sequences using a CRP rate parameter $\alpha=1$ and Dirichlet hyperparameter $\bm{\eta}=\mathbf{1}_V$. The fixed and variable-order Bayesian Markov (FBM, VBM) and Bayesian context tree (BCT) models are considered for comparison. Each variable-order model is specified a maximum depth $D=6$ before training. Table \ref{tab:combined-results} contains the results of fitting each model to the UniProt data. The fitted PBCT, VBM and BCT models each have maximum depth $D=4$. The PBCT achieves the best predictive performance in terms of log-loss with fewer parameters than the comparable models FBM-3, VBM and BCT.

\begin{figure}[t]
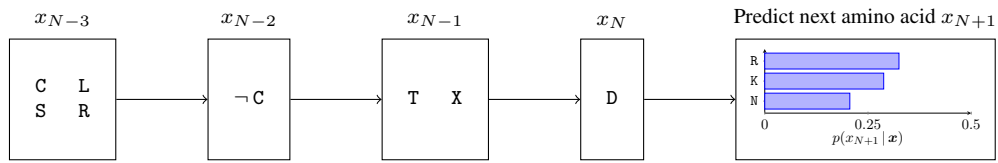

  \centering
   \ctikzfig{uniprot-interpretations}
    \caption{Example of protein motif discovery and prediction using the PBCT fitted to UniProt sequences. Here, the shorthand $\neg\,$\texttt{C} denotes all amino acids excluding \texttt{C}.}
     \label{fig:uniprot-interpret} 
\end{figure}

Figure~\ref{fig:uniprot-interpret} shows a learned order-4 context and the corresponding estimated probabilities for predicting the next amino acid. Aside from prediction, this example demonstrates an application of the PBCT to the discovery of protein sequence motifs \citep[see, for example,][]{Bailey2007}. Sequence motifs are short subsequences of amino acids used to characterise groups of proteins by functional or structural similarity. Fitting PBCTs to collections of protein sequences may lead to the discovery of new motifs, represented by learned contexts.

\section{Conclusion}
\label{sec:conclusion}
A novel inference method has been introduced to efficiently capture complex dependence structures and make predictions in discrete sequences. The proposed parsimonious Bayesian context tree model has been verified using a simulation study for a range of model configurations, and outperforms existing context tree models in real-world examples while reducing the parameter space. Key advantages of the PBCT model include scalability to large vocabularies and interpretability of results. PBCTs can be applied in practice for prediction tasks, anomaly detection and changepoint analysis, in addition to the protein motif discovery application discussed in Section~\ref{sec:uniprot}. An interesting extension of the inference method would involve updates of tree structures given streaming data. 

Markov chain Monte Carlo (MCMC) provides another method for inference of PBCT structure given data. MCMC via Gibbs sampling can be implemented by iteratively resampling from the posterior distributions of parent-child node structures. At each vocabulary clustering step, the results of agglomerative clustering can be used to initialise a Gibbs sampler. In such an implementation of MCMC, preliminary testing showed that the fitted tree structures rarely improved on the results using only RAC, Algorithm \ref{alg:agg-clust}.

\bigskip
\begin{center}
{\large\bf SUPPLEMENTARY MATERIAL}
\end{center}
The \textit{Python} library \texttt{pbct} contains the code used to fit parsimonious Bayesian context trees, available in the \textit{GitHub} repository at \url{https://github.com/daniyarghani/pbct}.

\bibliographystyle{apalike}
\bibliography{references.bib}

\end{document}